\documentclass[aps,amsmath,amssymb]{revtex4}

\usepackage{graphicx}
\usepackage{dcolumn}
\usepackage{bm}
\usepackage{amsmath}
\usepackage{epstopdf}
\usepackage{amsfonts}
\usepackage{amssymb}
\usepackage{epsfig}
\usepackage{tabularx}
\usepackage{color}
\usepackage{soul}

\usepackage[colorlinks=true,citecolor=blue,urlcolor=magenta,breaklinks]{hyperref}

\newcommand{\be}{\begin{equation}}
\newcommand{\en}{\end{equation}}
\newcommand{\bea}{\begin{eqnarray}}
\newcommand{\ena}{\end{eqnarray}}

\begin{document}

\title{Anisotropic Quark Stars with an Interacting Quark Equation of State within the Complexity Factor Formalism}

\author{
\'Angel Rinc\'on {${}^{a}$
\footnote{aerinconr@academicos.uta.cl}
} 
Grigoris Panotopoulos  {${}^{b}$
\footnote{grigorios.panotopoulos@ufrontera.cl}
}
Il{\'i}dio Lopes  {${}^{c}$
\footnote{ilidio.lopes@tecnico.ulisboa.pt}
}
}

\address{
${}^a$ Departamento de F{\'i}sica Aplicada, Universidad de Alicante, Campus de San Vicente del Raspeig, E-03690 Alicante, Spain;
\\
${}^b$ Departamento de Ciencias F\'isicas, Universidad de la Frontera,
Casilla 54-D, 4811186 Temuco, Chile.
\\
${}^c$ Centro de Astrof{\'i}sica e Gravita{\c c}{\~a}o-CENTRA, Instituto Superior T{\'e}cnico-IST, Universidade de Lisboa-UL, Av. Rovisco Pais, 1049-001 Lisboa, Portugal. 
}

\begin{abstract}
Within the framework of Einstein's General Relativity we study strange quark stars assuming an interacting equation-of-state. Taking into account the presence of anisotropies in a sphere made of ultra dense matter, we employ the formalism based on the complexity factor. We integrate the structure equations numerically imposing the appropriate conditions both at the center and at the surface of the stars, thus obtaining interior solutions describing hydrostatic equilibrium. Making use of well-established criteria, we demonstrate that the solutions obtained here are well behaved and realistic. A comparison with another, more conventional approach, is made as well. Our numerical results are summarized in a number of figures. .
\end{abstract}

\maketitle

\section{Introduction}\label{intro}
Unlike many other forms of matter, relativistic compact stars~\cite{Shapiro:1983du,Psaltis:2008bb,Lorimer:2008se,Zorotovic:2019uzl}, which are formed during the final stages of stellar evolution, are excellent cosmic probes to study the properties of matter under exceptionally extreme conditions.  Supernova explosions, either Type Ia or Type II, are the best known mechanisms to produce compact stars. Type Ia Supernovas occur when white dwarfs accrete matter from a companion in a binary (e.g.,~\cite{2020IAUS..357....1R}). Type II corresponds to the core collapse of massive stars; although the mass range is not well defined, the~current estimates predict a range from 8 to 30 solar masses (e.g.,~\cite{2021Natur.589...29B}). Although~the most likely stellar remnants of these events are neutron stars, they could also form many other kinds of compact stars.  Matter inside such objects is characterized by ultra-high mass/energy densities, for~which the usual description of astrophysical plasmas in terms of non-relativistic Newtonian fluids is inadequate. Therefore, those very dense compact objects are relativistic in nature, and~as such they 
are only properly described within the framework of Einstein’s General Relativity (GR)~\cite{Einstein:1915ca}.

Neutron stars (NSs) are exciting astronomical objects, since understanding and explaining their properties, as well as their observed complex phenomena, require bringing together several scientific disciplines and lines of research, such as nuclear particle physics, astrophysics and gravitational physics. As~NSs are the {densest} 
objects in the Universe after black holes, they can be used to probe, test and study properties of matter in an extreme gravitational regime, conditions that cannot be reproduced in earth-based experiments. 
For instance, supermassive star core collapses are known to trigger a complex network of physical processes
	that amplify magnetic fields up to tens of tera-tesla. This dramatic event involves neutrino emissions, convection~\cite{2015Natur.528..376M} and~magnetorotational instabilities~\cite{2015MNRAS.450.2153G}, accompanied by a turbulent dynamo affecting the synthesis of heavy nuclides by r- and other atomic process nuclei (e.g.,~\cite{2021Univ....7..487K}).    The~result is the formation of a compact star with a strong magnetic field in its interior, possibly with a high degree of anisotropy in the local equation of~state.   

Therefore, those objects constitute an excellent cosmic laboratory to explore and possibly constrain non-conventional physics, as well as alternative theories of~gravity.

Strange quark stars (QSs)~\cite{Alcock:1986hz,Alcock:1988re,Madsen:1998uh,Weber:2004kj,Yue:2006it,Leahy:2007we}, although
at the moment theorized and less standard astronomical objects than NSs, may be viewed as ultra-compact neutron stars. Since quark matter is by assumption absolutely stable, it may be the true ground state of hadronic matter~\cite{Witten:1984rs,Farhi:1984qu}. Therefore, this new class of relativistic compact objects has been proposed as an alternative to typical NSs. Though~as of today they remain hypothetical objects, quarks stars cannot conclusively be ruled out yet. 
In fact investigating their properties is well motivated, both from the theoretical and observational point of view. For example, there are some claims in the literature that there are currently some observed compact objects exhibiting peculiar features, such as small radii for instance, that cannot be explained by the usual hadronic equations-of-state used in NS studies, see e.g.,~\cite{Henderson:2007gu,Li:2011zzn,Aziz:2019rgf}, and~also Table~5 of~\cite{Weber:2004kj} and the references therein. The~present study is also relevant for the possible implications to understand the nature of compact stars. Recently, a~few authors suggested that strange matter could exist in the core of NS-hybrid stars~\cite{Benic:2014jia,Yazdizadeh_2022,EslamPanah:2018rfe}, while others claim such stars are almost indistinguishable from
NS~\cite{Jaikumar:2005ne}. In~addition to that, strange quark stars may explain some puzzling super-luminous supernovae~\cite{Ofek:2006vt,2009arXiv0911.5424O}, which occur in about one out of every 1000 supernovae explosions, and~which are more than 100 times more luminous than regular supernovae. One plausible explanation is that since quark stars are much more stable than NSs, they could explain the origin of the huge amount
of energy released in super-luminous supernovae. Many works have been recently proposed to validate their existence in
different astrophysical scenarios~\cite{Mukhopadhyay:2015xhs,Panotopoulos:2018ipq}.

In studies of compact objects, the authors usually focus their attention on relativistic stars made of isotropic matter,
such that $P_r=P_T$, where~$P_r$ is the radial pressure, and~$P_T$ is the tangential pressure. However, celestial bodies are not necessarily made of isotropic matter alone. As~a matter of fact, it is now known that under certain conditions the fluid representing matter content may become anisotropic. The~review article of Ruderman~\cite{Ruderman:1972aj} mentioned for the first time such a possibility: the author observed that interactions among relativistic particles in a very dense nuclear matter medium could generate a non-vanishing anisotropic factor. Studies on anisotropies in relativistic stars received a boost thanks to the subsequent work of~\cite{Bowers:1974tgi}. 
Indeed, anisotropies may arise in many different physical situations of a dense matter medium,
such as in the presence of strong magnetic fields~\cite{2017paln.book.....W,2022EPJC...82...74B}, in~phase transitions~\cite{Sokolov:1998svw}, in~pion condensation~\cite{Sawyer:1972cq},
slow rotation~\cite{1997PhR...286...53H}, viscosity-induced anisotropy
~\cite{1992Ap&SS.193..201B}, a~mixture of two fluids~\cite{1980PhRvD..22..807L}  
 or in the presence of type 3A super-fluid~\cite{Kippenhahn:2012qhp} (see also~\cite{MakANDHarko,Deb:2016lvi,Deb:2015vda} and the references therein for more recent works on the topic). In~those works, relativistic models of anisotropic quark stars were studied, and~the energy conditions were shown to be fulfilled. In~particular, in~\cite{MakANDHarko} an exact analytic solution was obtained, in~\cite{Deb:2016lvi} an attempt was made to find a 
singularity-free solution to Einstein’s field equations and~in~\cite{Deb:2015vda} the Homotopy Perturbation Method was employed, which is a tool that facilitates the ability to tackle Einstein’s field equations. Furthermore, alternative approaches have been considered to incorporate new anisotropic solutions to already known isotropic ones~\cite{Gabbanelli:2018bhs,Ovalle:2017fgl,Ovalle:2017wqi}.

In this work we propose to obtain interior solutions of relativistic stars, and in~particular, of quark stars made of anisotropic matter, within~the approach based on the complexity factor formalism. Originally, the~new definition of the complexity factor was only investigated from a mathematical point of view, see~\cite{Sharif:2018pgq,Sharif:2018efi,Abbas:2018cha,Herrera:2019cbx} and the references therein. However, the~real usefulness of 
such a definition becomes evident when we make use of it as a supplementary condition to close the set of structure equations describing the hydrostatic equilibrium of a self-gravitating spherical configuration. Moreover, the~complexity factor may be used as a self-consistent way to incorporate anisotropies, which are generated in the presence of ultra dense matter in relativistic objects, as~already mentioned~before.

The plan of our work in the present article is the following: after this introductory section, we review the description of relativistic stars in the framework of Einstein's theory in Section~\ref{GR}, while in the third section we present the vanishing complexity formalism. Next, in~Section~\ref{Dis} we show and discuss our numerical results, and~finally we finish our work with some concluding remarks in the last section. We work in geometrical units where $G=1=c$, and~we adopt the mostly negative metric signature in four space-time dimensions, namely $+,-,-,-$.

\section{Relativistic Stars in General~Relativity}\label{GR}

Let us start considering a static and spherically symmetric object in an anisotropic background bounded by a spherical surface $\Sigma$. Assuming a line element that is written in Schwarzschild-like coordinates, i.e.,~
\begin{equation}
ds^2=e^{\nu} dt^2 - e^{\lambda} dr^2 -
r^2 d\Omega^2.
\label{metric}
\end{equation}
The functions
  $\nu(r)$ and $\lambda(r)$ are the metric potentials and depend on the radial coordinate only, whereas $d\Omega^2\equiv \left( d\theta^2 + \sin^2\theta d\phi^2 \right)$ corresponds to the element of solid angle. We
will take: $x^0=t; \, x^1=r; \, x^2=\theta; \, x^3=\phi$. 
In the absence of a cosmological constant term, the~classical Einstein field equations acquire the simple form:
\begin{equation}
G^\nu_\mu=8\pi G T^\nu_\mu,
\label{Efeq}
\end{equation}
with $G$ being Newton's constant (taken to be unity in the following for simplicity). Now, in~the co-moving frame, the~physical matter content is an anisotropic fluid of energy density $\rho$,
radial pressure $P_r$, and~tangential pressure $P_\bot$, i.e.,~the covariant energy--momentum tensor in (local) Minkowski coordinates is then written as
$T^{\mu}_{\nu} =\{ \rho , P_r, P_\bot, P_\bot\}$ and the field equations can be written as:
\begin{eqnarray}
\rho &=& -\frac{1}{8\pi}\left[-\frac{1}{r^2}+e^{-\lambda}
\left(\frac{1}{r^2}-\frac{\lambda'}{r} \right)\right],
\label{fieq00}
\\
P_r &=& -\frac{1}{8\pi}\left[\frac{1}{r^2} - e^{-\lambda}
\left(\frac{1}{r^2}+\frac{\nu'}{r}\right)\right],
\label{fieq11}
\\
P_\bot &=& \frac{1}{32\pi}e^{-\lambda}
\left(2\nu''+\nu'^2 -
\lambda'\nu' + 2\frac{\nu' - \lambda'}{r}\right),
\label{fieq2233}
\end{eqnarray}
where the symbol prime ($\prime$) denotes the derivative in relation to $r$.

Combining
 Equations~(\ref{fieq00})--(\ref{fieq2233}), we obtain the hydrostatic equilibrium  equation, best known as the generalized Tolman--Opphenheimer--Volkoff equation, which takes the simplest form
\begin{equation}
-\frac{1}{2}\nu'\left( \rho + P_r\right)-P'_r+\frac{2}{r}\left(P_\bot-P_r\right)=0.\label{Prp}
\end{equation}
Such an equation can be understood as the balance between the following three forces: 
  (i)~gravitational, $F_g$, 
(ii) hydrostatic, $F_r$ and
(iii) anisotropic, $F_p$, 
defined according to the following expressions
\begin{equation}
F_g=-\frac{\nu'\left( \rho+ P_r\right)}{2},\;
\hspace{1cm}
F_r= -P'_r\; 
\hspace{1cm}
{\rm and}\;
\hspace{1cm}
F_p=\frac{2\Pi}{r}.
\label{Forces}
 \end{equation}  
At this point it should be mentioned that the anisotropic factor,
$\Pi$ (sometimes also referred to as $\Delta$), is defined as usual, i.e.,~ $ \Pi=P_\bot-P_r$.
{Accordingly}, Equation~(\ref{Prp}), now reads
\begin{equation}
F_g + F_r + F_p = 0.\label{Frp}
\end{equation}	
To clarify the situation, the~last equation implies that this compact star results from the equilibrium between these three different forces, as~is mentioned in Ref.~\cite{Prasad:2021eju}.  In~particular, if~$F_p=0$, we obtain the standard TOV equation.
Moreover, in~cases where  $P_\bot>P_r$  (or equivalently $\Pi>0$),  $F_p>0$ causes a repulsive force in Equation~(\ref{Frp}) that counteracts the attractive force given by the combination $F_g+F_r$.
On the contrary, if~we consider the case of $P_\bot<P_r$ (or $\Pi<0$),  $F_p<0$ is also an attractive force that adds to the other~ones.

To remove the $\nu'$-dependence in Equation~(\ref{Prp}), we
will take advantage of the following relation
\begin{equation}
\frac{1}{2}\nu' =  \frac{m + 4 \pi P_r r^3}{r \left(r - 2m\right)},
\label{nuprii}
\end{equation}
and we then rewrite the generalized TOV equation as
\begin{equation}
P'_r=-\frac{(m + 4 \pi P_r r^3)}{r \left(r - 2m\right)}\left( \rho + P_r\right)+\frac{2}{r}\left(P_\bot-P_r\right).\label{ntov}
\end{equation}
In addition, $m$ is the  mass function, obtained by:
\begin{equation}
R^3_{232}=1-e^{-\lambda}=\frac{2m}{r},
\label{rieman}
\end{equation}
or, 
\begin{equation}
m = 4\pi \int^{r}_{0} \tilde r^2\rho \  d\tilde r.
\label{m}
\end{equation}
The energy-momentum tensor can be rewritten as follows
\begin{equation}
T^{\mu}_{\nu}=\rho u^{\mu}u_{\nu}-  P
h^{\mu}_{\nu}+\Pi ^{\mu}_{\nu}.
\label{24'}
\end{equation}
From the expression of $T_{\mu}^{\nu}$, we have to point out that the four-velocity is taken to be 
$u^{\mu} = (e^{-\frac{\nu}{2}},0,0,0)$, and~the four-acceleration is $a^\alpha=u^\alpha_{;\beta}u^\beta$, whose any non-vanishing component is $a_1 = -\nu^{\prime}/2$. 
Subsequently, the~set $\{ \Pi^{\mu}_{\nu}, \Pi, h^{\mu}_{\nu}, s^{\mu}, P \}$ is defined as
\begin{eqnarray}
\Pi^{\mu}_{\nu} &=& \Pi\bigg(s^{\mu}s_{\nu}+\frac{1}{3}h^{\mu}_{\nu}\bigg)
\\
\Pi &=& P_{\bot}-P_r \label{Delta}
\\
h^\mu_\nu &=& \delta^\mu_\nu-u^\mu u_\nu
\\
s^{\mu} &=& (0,e^{-\frac{\lambda}{2}},0,0)\label{ese}
\\
P & \equiv & \frac{1}{3}\Bigl( P_{r}+2P_{\bot} \Bigl)
\end{eqnarray}
with the  properties
$s^{\mu}u_{\mu}=0$ and
$s^{\mu}s_{\mu}=-1$.
For the exterior solution, we match the problem with Schwarzschild space-time, i.e.,
\begin{equation}
ds^2= \left(1-\frac{2M}{r}\right) dt^2 - \left(1-\frac{2M}{r}\right)^{-1} dr^2 -
r^2  d\Omega^2.
\label{Vaidya}
\end{equation}
We have to complete the problem with the appropriate matching conditions on the surface
$r=R=\text{constant}$, with~$R$ being the radius of the star. Thus, by demanding continuity of the first and the second fundamental forms across that surface, we have
\begin{eqnarray}
e^{\nu_\Sigma} &=& 1-\frac{2M}{R},
\label{enusigma}
\\
e^{-\lambda_\Sigma} &=& 1-\frac{2M}{R},
\label{elambdasigma}
\\
\left[P_r\right]_\Sigma &=& 0,
\label{PQ}
\end{eqnarray}
 where $M$ is the mass of the star, while the subscript $\Sigma$ indicates that the quantity is evaluated on the boundary surface $\Sigma$.
To finish, let us mention that the last three equations are the necessary and sufficient conditions for a smooth 
matching of the interior and exterior solutions, (\ref{metric}) and (\ref{Vaidya}) respectively, on~the surface $\Sigma$.

\section{Vanishing Complexity Factor~Formalism}\label{VCF}

The present section summarizes the main ingredient behind the complexity factor formalism and its importance in astrophysical scenarios. To~start, we should mention the original paper of Luis Herrera (L.H.)~\cite{Herrera:2018bww}, in which the study of static anisotropic self-gravitating objects was performed. In~that paper, the~author introduced a modern definition of complexity and was motivated mainly by two concrete problems present in old descriptions: 
(i) the probability distribution is replaced by the energy density of the fluid distribution~\cite{Sanudo:2008bu}, and~(ii) a complete inclusion of the components of the energy density fluid is missing, and~the unique inclusion is the energy density of the fluid (ignoring possible additional contributions such as pressure).
These two critical points are, in~principle, the~primary motivation used by L.H. to modify the more traditional complexity~definition. 

Albeit the standard definition of complexity was first introduced as a mathematical treatment only (see~\cite{Sharif:2018pgq,Sharif:2018efi,Abbas:2018cha,Herrera:2019cbx} and the references therein), its importance becomes evident when we recognize that such a condition can be used as an alternative route to bypass a standard ingredient in stellar interiors: the election of certain profiles of the system. Thus, the~complexity factor formalism, or, more precisely, its simplified version, namely, when the complexity factor is zero, allows us to close the system and obtain novel and non-trivial spherically symmetric solutions unknown up to now. Thus, the~vanishing complexity formalism offers a parallel way to deal with anisotropies in relativistic stars (see for instance~\cite{Arias:2022qrm,Andrade:2021flq,Contreras:2022vec,Bargueno:2022yob,Contreras:2021xkf}).

Now, let us move to the basic details behind the method. The~essence of the complexity factor formalism emerges when we check the orthogonal splitting of the Riemann tensor for static self-gravitating fluids with spherical symmetry (please, for~a detailed explanation, see~\cite{Herrera:2018bww} and also~\cite{Gomez-Lobo:2007mbg}). The~(orthogonal) decomposition of the Riemann tensor is usually tedious labor and was well-studied in the original paper, thus we will avoid repetitions. We will take advantage of some helpful definitions:
\begin{eqnarray}
Y_{\alpha \beta} &=& R_{\alpha \gamma \beta \delta}u^{\gamma}u^{\delta}, \label{electric} 
\\    
Z_{\alpha \beta} &=& R^{*}_{\alpha \gamma \beta
\delta}u^{\gamma}u^{\delta} = \frac{1}{2}\eta_{\alpha \gamma
\epsilon \mu} R^{\epsilon \mu}_{\quad \beta \delta} u^{\gamma}
u^{\delta}, \label{magnetic} 
\\
X_{\alpha \beta} &=& R^{*}_{\alpha \gamma \beta \delta}u^{\gamma}u^{\delta}=
\frac{1}{2}\eta_{\alpha \gamma}^{\quad \epsilon \mu} R^{*}_{\epsilon
\mu \beta \delta} u^{\gamma}
u^{\delta}, \label{magneticbis}
\end{eqnarray}
where the symbol $*$ is the dual tensor, i.e.,~
\begin{equation}
   R^{*}_{\alpha \beta \gamma \delta}=\frac{1}{2}\eta_{\epsilon \mu \gamma \delta}R_{\alpha \beta}^{\quad \epsilon \mu} 
\end{equation}
and $\eta_{\epsilon \mu \gamma \delta}$ is the well-known Levi--Civita tensor.
Utilizing the  decomposition of the Riemann tensor,  
the set of tensors
$\{Y_{\alpha \beta}, Z_{\alpha \beta}, X_{\alpha \beta}\}$ 
can be written in a convenient way, in~terms of the physical variables, namely, 
\begin{eqnarray}
Y_{\alpha\beta} &=& \frac{4\pi}{3}(\rho +3
P)h_{\alpha\beta}+4\pi \Pi_{\alpha\beta}+E_{\alpha\beta},\label{Y}
\\
Z_{\alpha\beta} &=& 0,\label{Z}
\\
X_{\alpha\beta} &=& \frac{8\pi}{3} \rho
h_{\alpha\beta}+4\pi
 \Pi_{\alpha\beta}-E_{\alpha\beta}.\label{X}
\end{eqnarray}
Also, the~tensor $E_{\alpha \beta}$ (defined as $E_{\alpha \beta}=C_{\alpha \gamma \beta
\delta}u^{\gamma}u^{\delta}$) can be written as
\begin{equation}
E_{\alpha \beta}=E \bigg(s_\alpha s_\beta+\frac{1}{3}h_{\alpha \beta}\bigg),
\label{52bisx}
\end{equation}
where $E$ is precisely
\begin{equation}
E=-\frac{e^{-\lambda}}{4}\left[ \nu ^{\prime \prime} + \frac{{\nu
^{\prime}}^2-\lambda ^{\prime} \nu ^{\prime}}{2} -  \frac{\nu
^{\prime}-\lambda ^{\prime}}{r}+\frac{2(1-e^{\lambda})}{r^2}\right],
\label{defE}
\end{equation}
satisfying the following properties:
\begin{eqnarray}
 E^\alpha_{\,\,\alpha}=0,\quad E_{\alpha\gamma}=
 E_{(\alpha\gamma)},\quad E_{\alpha\gamma}u^\gamma=0.
  \label{propE}
 \end{eqnarray} 
At this point it is essential to mention that the tensors $\{ Y_{\alpha\beta}, Z_{\alpha\beta}, X_{\alpha\beta} \}$ can be expressed in terms of alternative scalar functions, as~was previously discussed in~\cite{Herrera:2009zp}.
Now, consider the following tensors: $X_{\alpha \beta}$ and $Y_{\alpha \beta}$ in the static case. Based on that, it is possible to define the so-called structure scalars  $X_T, X_{TF}, Y_T, Y_{TF}$, in~terms of the physical variables, as~follows:
\begin{align}
X_T     &= 8\pi  \rho ,  \label{esnIII} 
\\
X_{TF}  &= \frac{4\pi}{r^3} \int^r_0{\tilde r^3 \rho ' d\tilde r} \label{defXTFbis}, 
\\
Y_T     &= 4\pi( \rho  + 3 P_r-2\Pi) \label{esnV}, 
\\
Y_{TF}  &= 8\pi \Pi- \frac{4\pi}{r^3} \int^r_0{\tilde r^3 \rho' d\tilde r} \label{defYTFbis}.
\end{align}
From Equations~\eqref{defXTFbis}--\eqref{defYTFbis}, the~local anisotropy of pressure is  obtained utilizing  $X_{TF}$ and $Y_{TF}$ with help of the following relation:
\begin{equation}
8\pi \Pi = X_{TF} + Y_{TF} \label{defanisxy}.
\end{equation}
The vanishing complexity condition, $Y_{TF}=0$, implies the following relation between the energy density and the 
anisotropic factor
\begin{equation}
\Pi(r) = \frac{1}{2r^3} \: \int^r_0{\tilde r^3 \rho'(\tilde r) d\tilde r}.
\end{equation}
Thus, for~a given density profile $\rho(r)$, we can compute the anisotropic factor for a self-gravitating system. Notice that the classical formalism, i.e.,~selecting a suitable form of the anisotropic factor by hand (among other alternatives), has also been significantly investigated over the years (see, for~instance,~\cite{Panotopoulos:2018joc,Panotopoulos:2018ipq,Moraes:2021lhh,Gabbanelli:2018bhs,Panotopoulos:2019wsy,Lopes:2019psm,Panotopoulos:2019zxv,Abellan:2020jjl,Panotopoulos:2020zqa,Bhar:2020ukr,Panotopoulos:2020kgl,Panotopoulos:2021obe,Panotopoulos:2021dtu} and the references therein).
As the complexity factor formalism in the context of relativistic stars is so helpful, we will follow this idea and contrast the results obtained following such an approach with solutions coming from the canonical method, i.e.,~taking by hand the anisotropic factor, for~instance.

\section{Results and~Discussion}\label{Dis}

In the present paper we have investigated within GR anisotropic stars made of quark matter in light of the vanishing complexity formalism. In~particular, we compute numerically interior solutions of realistic spherical configurations of anisotropic matter, and~we compare our solution against a more conventional approach, i.e., assuming an ansatz for the anisotropic factor by hand. For~quark matter we adopt the interacting equation-of-state, $P_r(\rho)$, given by~\cite{Becerra-Vergara:2019uzm,Panotopoulos:2021cxu}
\begin{eqnarray} \label{Prad1}
&& P_r = \dfrac{1}{3}\left(\rho-4B\right)-\dfrac{m_{s}^{2}}{3\pi}\sqrt{\dfrac{\rho-B}{a_4}} \nonumber\\
&& +\dfrac{m_{s}^{4}}{12\pi^{2}}\left[1-\dfrac{1}{a_4}+3\ln\left(\dfrac{8\pi}{3m_{s}^{2}}\sqrt{\dfrac{\rho-B}{a_4}}\right)\right],
\end{eqnarray}
where $P_r$ is the radial pressure, and~$\rho$ is the energy density of homogeneously distributed quark matter (also to $\mathcal{O}$ $(m_s^4)$ in the Bag model). For~the purpose of the present analysis, following Beringer \textit{et al}~\cite{ParticleDataGroup:2012pjm}, we take the strange quark mass ($m_{s}$) to be $100 \,{\rm MeV}$, while the accepted values of the bag constant, $B$, lie within the range $57 \leq B \leq 92$~MeV/fm$^3$~\cite{FiorellaBurgio:2018dga,Blaschke:2018mqw}. Also, the~parameter $a_4$ comes from QCD corrections on the pressure of the quark-free Fermi sea, and~it is directly related with the mass-radius relations of quark stars. Note that as far as the EoS is concerned, only the radial pressure is relevant. The~tangential pressure has already been defined in the Introduction as well as after Equation~(2).

\begin{figure}[ht!]

\includegraphics[width=0.32\textwidth]{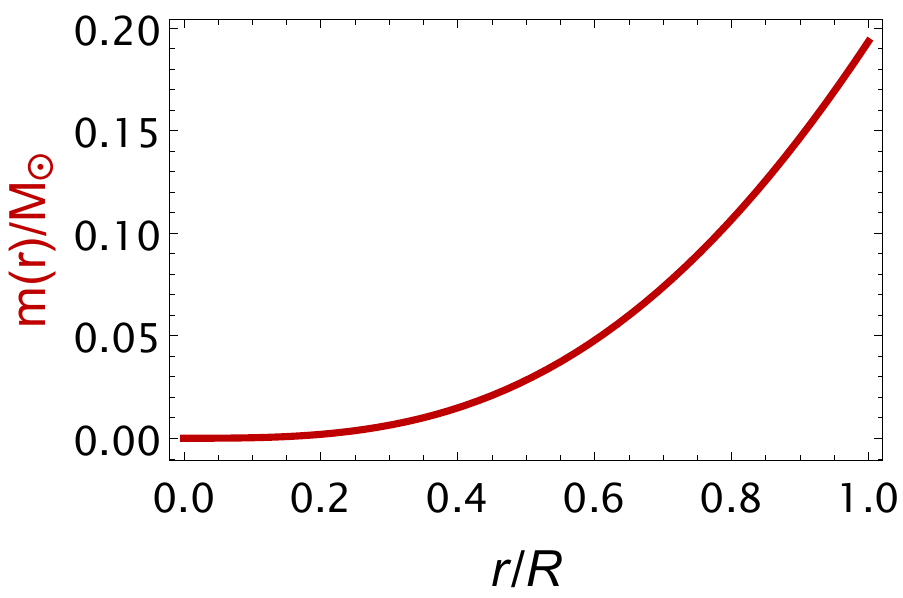} 
\includegraphics[width=0.32\textwidth]{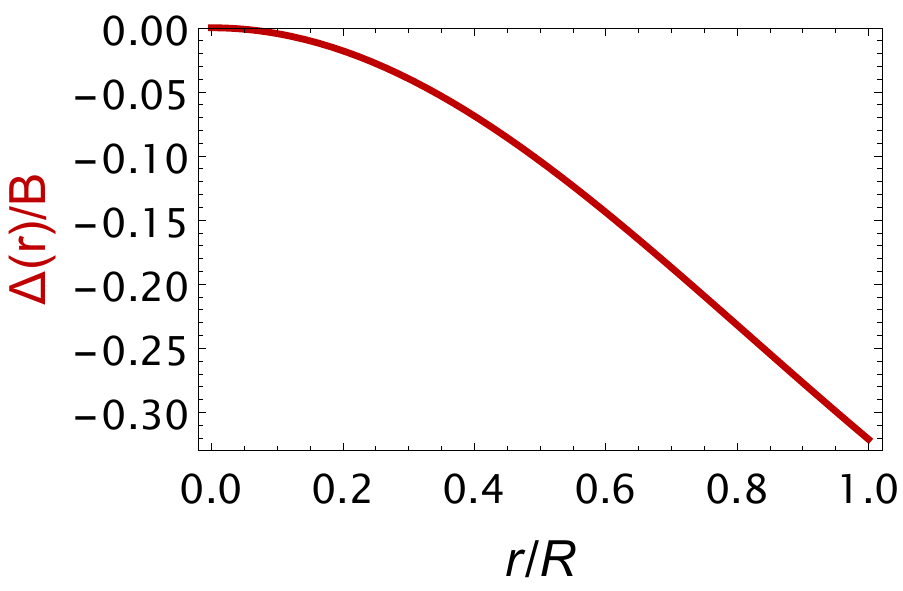}  
\includegraphics[width=0.32\textwidth]{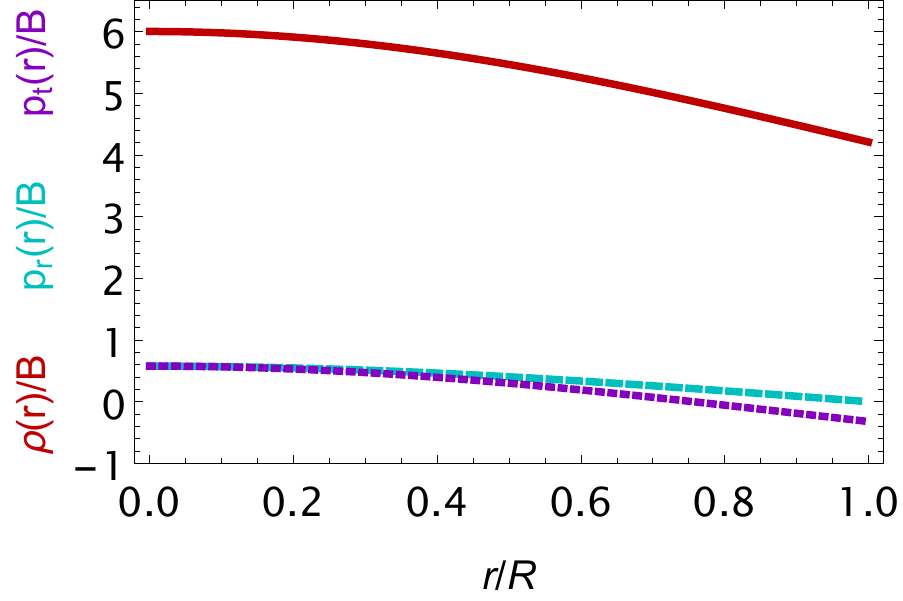} 
\caption{Anisotropic SQ stars within {the} 
	complexity factor: mass function in solar masses (\textbf{left} panel), anisotropic factor (\textbf{middle} panel), and~energy density and pressures (\textbf{right} panel) versus the radial coordinate throughout the star.} 
\label{fig:1} 	
\end{figure}

\begin{figure}[ht!]

\includegraphics[width=0.48\textwidth]{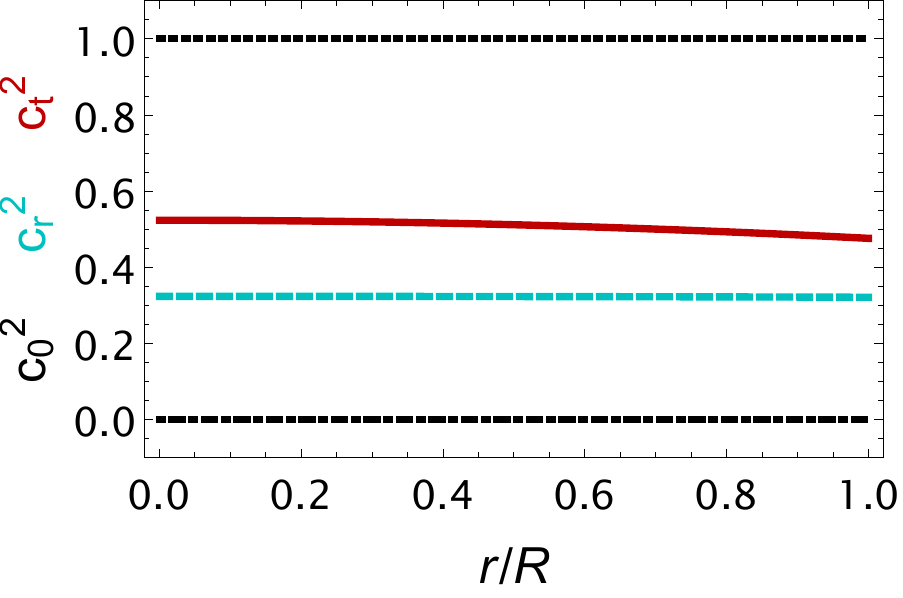} 
\includegraphics[width=0.48\textwidth]{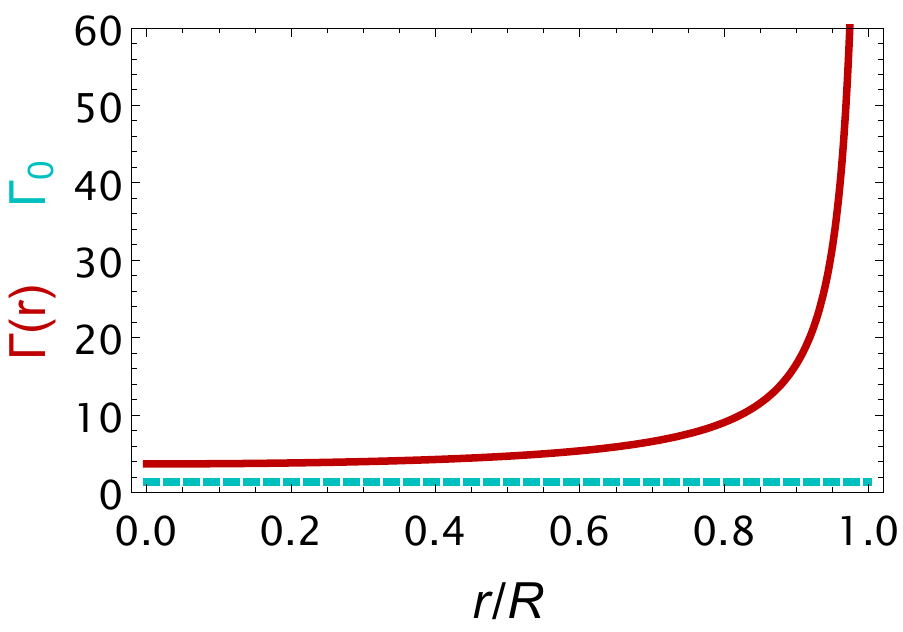} 
\caption{Anisotropic SQ stars within {the} 
	complexity factor: speed of sounds (\textbf{left} panel) and relativistic adiabatic index (\textbf{right} panel) versus the radial coordinate throughout the star. The horizontal straight line corresponds to the value $\Gamma_0=4/3$.}

\label{fig:2} 	
\end{figure}

\begin{figure}[ht]

\includegraphics[width=0.48\textwidth]{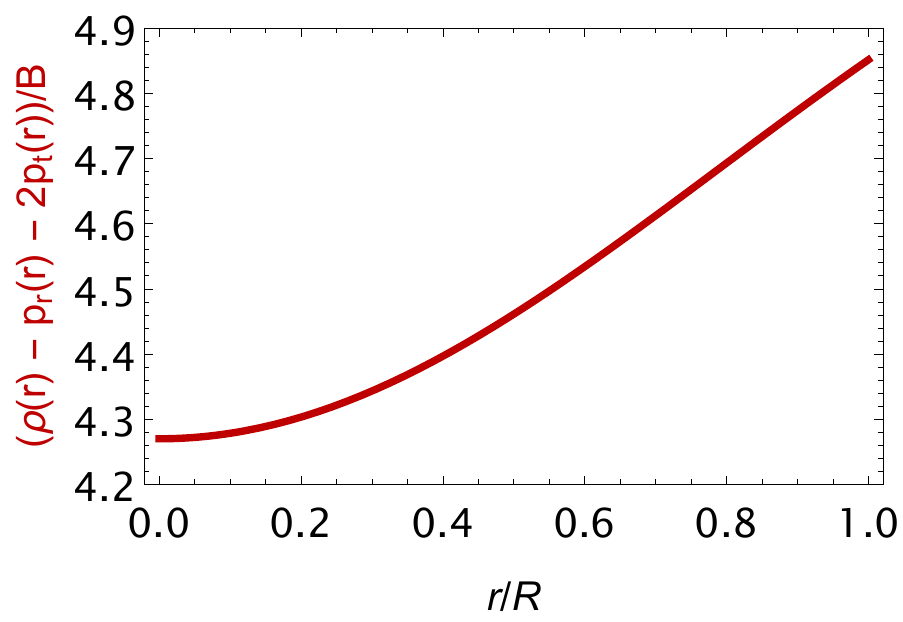} \
\includegraphics[width=0.48\textwidth]{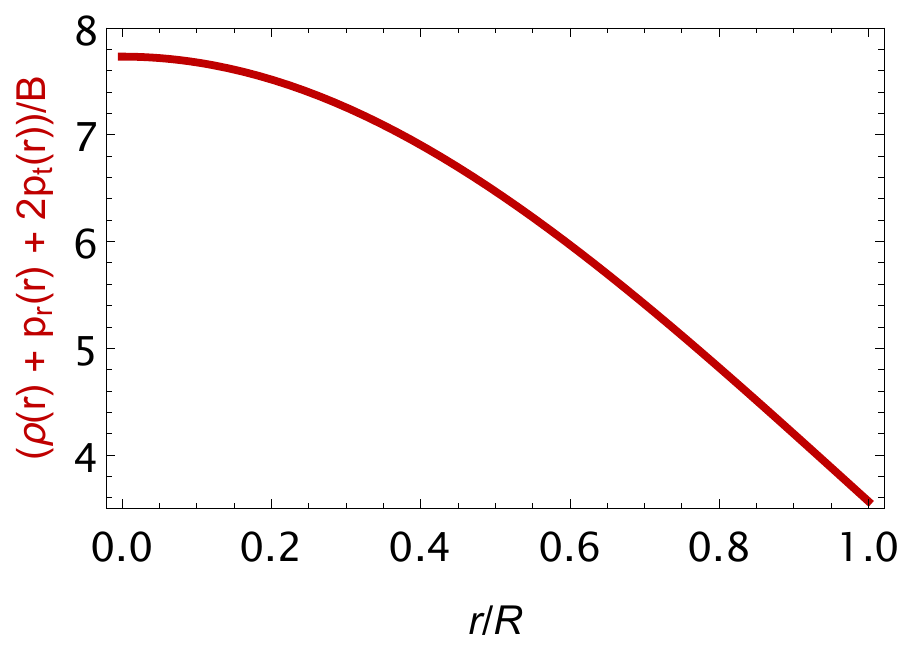} 
\caption{
Anisotropic QS stars within {the} 
complexity factor: energy conditions versus the radial coordinate throughout the star.
}
\label{fig:3} 	
\end{figure}

\begin{figure}[ht!]

\includegraphics[width=0.32\textwidth]{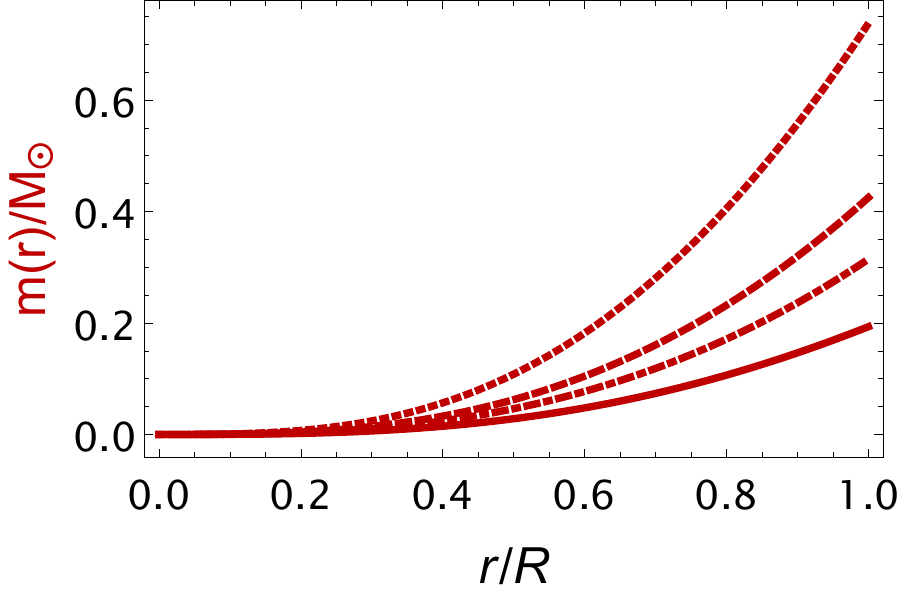} \
\includegraphics[width=0.32\textwidth]{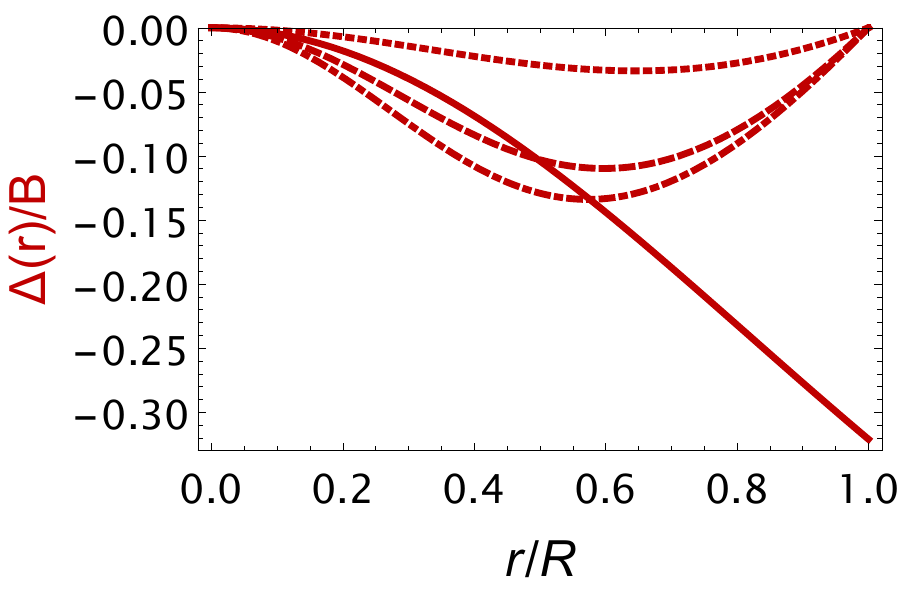} \
\includegraphics[width=0.32\textwidth]{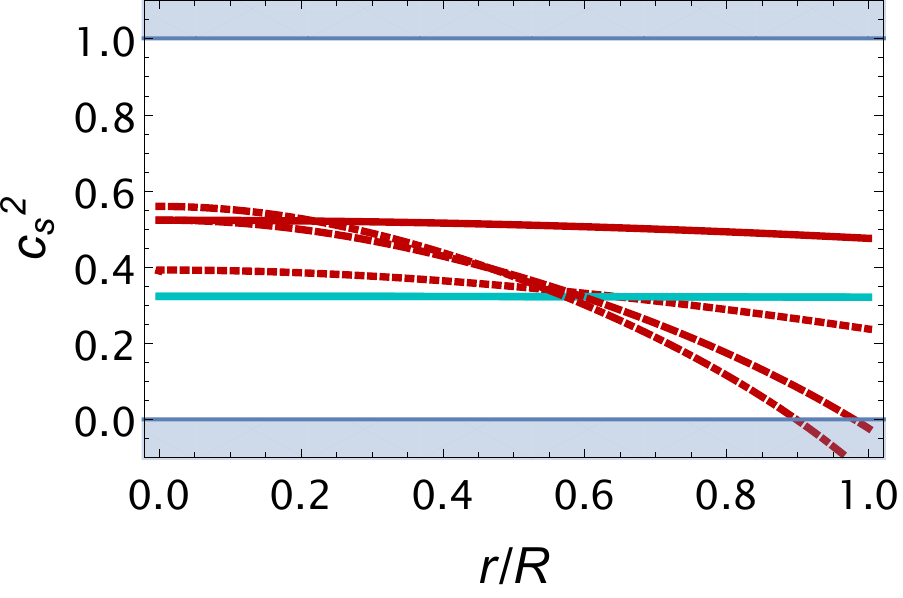} 
\caption{
Anisotropic QS stars considering a more standard approach and assuming 
  (i) $\kappa=-1$ (dotted red line),
 (ii) $\kappa=-6$ (dashed red line),
(iii) $\kappa=-10$ (dot-dashed red line), 
for:
 mass function in solar masses (\textbf{left} panel), anisotropic factor (\textbf{middle} panel), and~sound speeds (\textbf{right} panel) versus the radial coordinate throughout the star.
In particular, for~the right-hand panel, we have plotted tangential (red color) and radial (cyan color) $c_s^2$, respectively.
For comparison, we added our solution utilizing the complexity formalism (solid red line).
}
\label{fig:4} 	
\end{figure}

After the numerical computations, we display in a number of figures, several quantities of interest.
In Figures~\ref{fig:1}--\ref{fig:4}, on~the x-axis we put the normalized radial coordinate $r/R$, while on the y-axis we put the quantities of interest, such as the mass of the star in solar masses, dimensionless quantities (for instance, the~relativistic adiabatic index), and~normalized quantities (e.g., the pressures and the energy density as well as the anisotropic factor in units of the bag constant, B). The~curves exhibit the usual behaviour observed in interior solutions describing hydrostatic equilibrium. In~particular, the~mass function increases starting from 0 at the centre, while its mass at $r=R$ corresponds to the star's mass. The~energy density and the pressures start from their central value and monotonically decrease until the radial pressure vanishes at the star's surface. In~contrast, the~energy density acquires a surface value.
The energy density and the pressures remain positive throughout the star, while the energy density always remains larger than the pressures, such that~the energy conditions are fulfilled. Moreover, both sound speeds vary slowly with the radial coordinate, taking values within the range 0--1 throughout the star, so causality is not violated. The~anisotropic factor turns out to be negative, which is always the case within the vanishing complexity formalism. Finally, the~relativistic adiabatic index increases monotonically and rapidly acquires large values as we approach the surface of the star. At~the same time, it always remains larger than the Newtonian value $\Gamma_0=4/3$.
In particular, we notice that: 
(i) the mass function increase, the~anisotropic factor decrease and the energy density and pressures decrease throughout the star, (ii) the speed of sound, radial and tangential, increase and decrease, respectively, and~both are lower {than} 
$c_0^2 \equiv 1$, the~relativistic adiabatic index, $\Gamma(r)$, increase and it is always higher than $\Gamma_0 \equiv 4/3$, 
(iii) the corresponding energy conditions are also satisfied. Thus, in~light of the these numerical results, we can confirm that the complexity factor formalism is a solid approach to obtain well-defined solutions in the context of compact~stars. 

As a supplementary independent check, we have obtained, numerically again, interior solutions using a more standard approach, i.e.,~adding external constraints to close the system of differential equations. As~a toy model, we have considered an anisotropic factor, $\Pi(r)$, as~follows~\cite{Silva:2014fca,Folomeev:2015aua,Cattoen:2005he,Horvat:2010xf,Arbanil:2021ahh}
\begin{equation}	
\Pi(r) = \kappa \: P_r(r) \: [ 1-e^{-\lambda(r)} ]= \kappa \: P_r(r) \: \frac{2m}{r},
\end{equation}
characterized by a dimensionless parameter, $\kappa$, which encodes the strength of the anisotropy. Also, we consider several values from $\kappa=-1$ to $\kappa=-10$.


\unskip

 
\unskip


\unskip


Our main results may be summarized as follows:  the conventional method predicts more massive objects in comparison to the stars within the vanishing complexity formalism. The~mass of the object decreases with the anisotropic factor, approaching the mass corresponding to the method based on vanishing complexity.
Eventually, causality is violated as the tangential speed of sound becomes negative, and~therefore the solution is no longer~realistic.

\begin{figure}[ht!]

\includegraphics[width=0.75\textwidth]{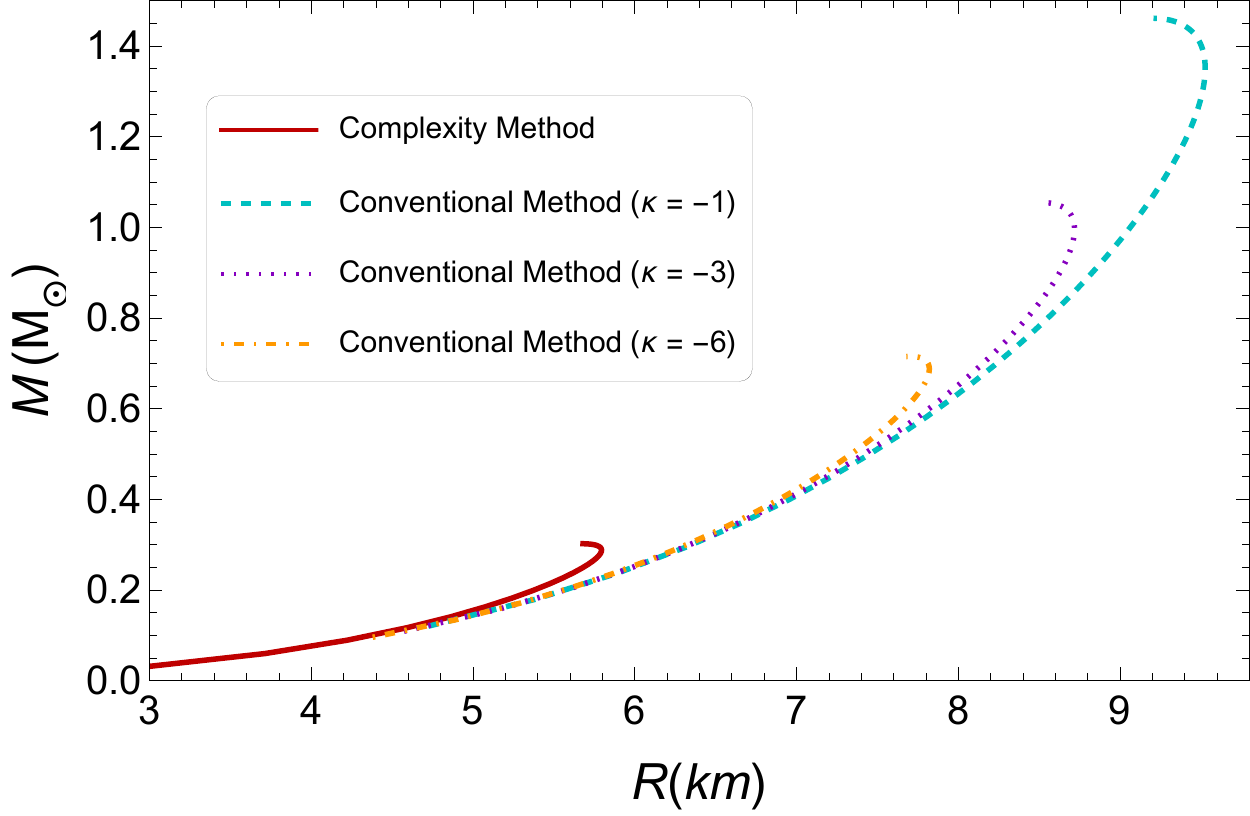} 
\caption{Mass-radius profiles obtained by two alternative methods:
  (i) the vanishing complexity method (solid red line), and~(ii) the conventional method for: $\kappa=-1$ (cyan dashed line), $\kappa = -3$ (dashed purple line), and~$\kappa = -6$ (dot-dashed orange line). 
}
\label{fig:7} 	
\end{figure}

Finally,  in Figure \eqref{fig:7},  we show the mass-to-radius relationship for anisotropic stars for both approaches.
First we study anisotropies within the vanishing complexity formalism, and~we obtain the curve in red. Next, when we consider the conventional method, since now the ansatz for the anisotropic factor is characterized by a continuous parameter, we can observe the impact of that parameter on the profiles, corresponding to the other 3 curves in the figure. Increasing the anisotropy, the~profile is gradually shifted towards the one corresponding to complexity. But~at some point causality is violated, and~therefore the solution is not realistic/viable any more. That is precisely the point where we must stop. The~last allowed profile remains quite far away from the one obtained within~complexity.

Before we conclude our work, a final comment is in order. In~the present article our main goal was to study the
implications of the formalism based on vanishing complexity factor, and~compare to a more standard approach. It would
be interesting, however, to~study the stability of anisotropic stars as well, analyzing radial oscillation modes and computing the corresponding frequencies. As~this investigation lies beyond the scope of the present work, we propose to postpone it
for the time being, and~we hope to be able to do so in the future.

\section{Conclusions}\label{Conc}

To summarize our work, in~the present article we have obtained interior solutions of strange quark stars made of anisotropic matter. Assuming an interacting equation-of-state, we have taken into account the presence of anisotropies generated by ultra dense quark matter. The~anisotropic factor has been introduced employing the formalism based on the complexity factor, and~the structure equations have been integrated numerically. The~solutions have been shown to be well-behaved and realistic. Moreover, we have made a comparison with another more conventional approach, where the form of the anisotropic factor is introduced by hand. Finally, we have obtained the mass-to-radius relationships within both methods, namely (i) employing the complexity formalism (where there are no free parameters), and~(ii) adopting a certain form for the anisotropic factor (used in previous works, characterized by a free parameter). Our results show that as the anisotropy {inceases, thus varying} 
the free parameter, the~profile is shifted towards the one corresponding to the complexity factor~formalism.

\section{Acknowledgments}

A.R. is funded by the Mar{\'i}a Zambrano contract ZAMBRANO 21-25 (Spain). I.L. thanks the Funda\c c\~ao 
para a Ci\^encia e Tecnologia (FCT), Portugal, for~the financial support to the Center for Astrophysics and 
Gravitation (CENTRA/IST/ULisboa)  through the Grant Project No.~UIDB/00099/2020  and Grant No.~PTDC/FIS-AST/28920/2017.



\begin{thebibliography}{999}
	
\bibitem[Shapiro and Teukolsky(1983)]{Shapiro:1983du}
Shapiro, S.L.; Teukolsky, S.A.
\newblock {\em Black Holes, White Dwarfs, and Neutron Stars: The Physics of
Compact Objects}; John Wiley \& Sons: Hoboken, NJ, USA, 1983.

\bibitem[Psaltis(2008)]{Psaltis:2008bb}
Psaltis, D.
\newblock {Probes and Tests of Strong-Field Gravity with Observations in the
Electromagnetic Spectrum}.
\newblock {\em Living Rev. Rel.} {\bf 2008}, {\em 11},~9.
\newblock {{https://doi.org/10.12942/lrr-2008-9}}.
\bibitem[Lorimer(2008)]{Lorimer:2008se}
Lorimer, D.R.
\newblock {Binary and Millisecond Pulsars}.
\newblock {\em Living Rev. Rel.} {\bf 2008}, {\em 11},~8.
\newblock {{https://doi.org/10.12942/lrr-2008-8}}.

\bibitem[Zorotovic and Schreiber(2020)]{Zorotovic:2019uzl}
Zorotovic, M.; Schreiber, M.R.
\newblock {Cataclysmic variable evolution and the white dwarf mass problem: A
Review}.
\newblock {\em Adv. Space Res.} {\bf 2020}, {\em 66},~1080--1089.
\newblock {{https://doi.org/10.1016/j.asr.2019.08.044}}.

\bibitem[{Ruiter}(2020)]{2020IAUS..357....1R}
{Ruiter}, A.J.
\newblock {Type Ia supernova sub-classes and progenitor origin}.
\newblock {\em IAU Symp.} {\bf 2020}, {\em 357},~1--15.
\newblock {{https://doi.org/10.1017/\linebreak S1743921320000587}}.

\bibitem[{Burrows} and {Vartanyan}(2021)]{2021Natur.589...29B}
{Burrows}, A.; {Vartanyan}, D.
\newblock {Core-collapse supernova explosion theory}.
\newblock {\em Nature} {\bf 2021}, {\em 589},~29--39.
\newblock {{https://doi.org/10.1038/s41586-020-03059-w}}.

\bibitem[Einstein(1915)]{Einstein:1915ca}
Einstein, A.
\newblock {The Field Equations of Gravitation}.
\newblock {\em Sitzungsber. Preuss. Akad. Wiss. Berl. (Math. Phys.)} {\bf
1915}, {\em 1915},~844--847.

\bibitem[{M{\"o}sta} \em{et~al.}(2015){M{\"o}sta}, {Ott}, {Radice}, {Roberts}.
{Schnetter}, and {Haas}]{2015Natur.528..376M}
{M{\"o}sta}, P.; {Ott}, C.D.; {Radice}, D.; {Roberts}, L.F.; {Schnetter}, E.;
{Haas}, R.
\newblock {A large-scale dynamo and magnetoturbulence in rapidly rotating
core-collapse supernovae}.
\newblock {\em Nature} {\bf 2015}, {\em 528},~376--379.
\newblock {{https://doi.org/10.1038/nature15755}}.

\bibitem[{Guilet} and {M{\"u}ller}(2015)]{2015MNRAS.450.2153G}
{Guilet}, J.; {M{\"u}ller}, E.
\newblock {Numerical simulations of the magnetorotational instability in
protoneutron stars---I. Influence of buoyancy}.
\newblock {\em Mon. Not. R. Astron. Soc.} {\bf 2015}, {\em 450},~2153--2171.
\newblock {{https://doi.org/10.1093/mnras/stv727}}.

\bibitem[{Kondratyev}(2021)]{2021Univ....7..487K}
{Kondratyev}, V.N.
\newblock {R-Process with Magnetized Nuclei at Dynamo-Explosive Supernovae and
Neutron Star Mergers}.
\newblock {\em Universe} {\bf 2021}, {\em 7},~487.
\newblock {{https://doi.org/10.3390/universe7120487}}.

\bibitem[Alcock \em{et~al.}(1986)Alcock, Farhi, and Olinto]{Alcock:1986hz}
Alcock, C.; Farhi, E.; Olinto, A.
\newblock {Strange stars}.
\newblock {\em Astrophys. J.} {\bf 1986}, {\em 310},~261--272.
\newblock {{https://doi.org/10.1086/164679}}.

\bibitem[Alcock and Olinto(1988)]{Alcock:1988re}
Alcock, C.; Olinto, A.
\newblock {Exotic Phases of Hadronic Matter and their Astrophysical
Application}.
\newblock {\em Ann. Rev. Nucl. Part. Sci.} {\bf 1988}, {\em 38},~161--184.
\newblock {{https://doi.org/10.1146/annurev.ns.38.120188.001113}}.

\bibitem[Madsen(1999)]{Madsen:1998uh}
{Madsen, J.
\newblock {Physics and astrophysics of strange quark matter}.
\newblock {\em Lect. Notes Phys.} {\bf 1999}, {\em 516},~162--203.
\newblock {{https://doi.org/10.1007/\linebreak BFb0107314}}.}

\bibitem[Weber(2005)]{Weber:2004kj}
Weber, F.
\newblock {Strange quark matter and compact stars}.
\newblock {\em Prog. Part. Nucl. Phys.} {\bf 2005}, {\em 54},~193--288.
\newblock {{https://doi.org/10.1016/\linebreak j.ppnp.2004.07.001}}.

\bibitem[Yue \em{et~al.}(2006)Yue, Cui, and Xu]{Yue:2006it}
Yue, Y.L.; Cui, X.H.; Xu, R.X.
\newblock {Is psr b0943+10 a low-mass quark star?}
\newblock {\em Astrophys. J. Lett.} {\bf 2006}, {\em 649},~L95--L98.
\newblock {{https://doi.org/10.1086/\linebreak508421}}.

\bibitem[Leahy and Ouyed(2008)]{Leahy:2007we}
Leahy, D.; Ouyed, R.
\newblock {Supernova SN2006gy as a first ever Quark Nova?}
\newblock {\em Mon. Not. R. Astron. Soc.} {\bf 2008}, {\em 387},~1193.
\newblock {{https://doi.org/10.1111/j.1365-2966.2008.13312.x}}.

\bibitem[Witten(1984)]{Witten:1984rs}
Witten, E.
\newblock {Cosmic Separation of Phases}.
\newblock {\em Phys. Rev. D} {\bf 1984}, {\em 30},~272--285.
\newblock {{https://doi.org/10.1103/PhysRevD.30.272}}.

\bibitem[Farhi and Jaffe(1984)]{Farhi:1984qu}
Farhi, E.; Jaffe, R.L.
\newblock {Strange Matter}.
\newblock {\em Phys. Rev. D} {\bf 1984}, {\em 30},~2379.
\newblock {{https://doi.org/10.1103/PhysRevD.30.2379}}.

\bibitem[Henderson and Page(2007)]{Henderson:2007gu}
Henderson, J.A.; Page, D.
\newblock {RX J1856.5-3754 as a possible Strange Star candidate}.
\newblock {\em Astrophys. Space Sci.} {\bf 2007}, {\em 308},~513--517.
\newblock {{https://doi.org/10.1007/s10509-007-9329-7}}.

\bibitem[Li \em{et~al.}(2011)Li, Peng, and Lu]{Li:2011zzn}
Li, A.; Peng, G.X.; Lu, J.F.
\newblock {Strange star candidates revised within a quark model with chiral
mass scaling}.
\newblock {\em Res. Astron. Astrophys.} {\bf 2011}, {\em 11},~482--490.
\newblock {{https://doi.org/10.1088/1674-4527/11/4/010}}.

\bibitem[Aziz \em{et~al.}(2019)Aziz, Ray, Rahaman, Khlopov, and
Guha]{Aziz:2019rgf}
Aziz, A.; Ray, S.; Rahaman, F.; Khlopov, M.; Guha, B.K.
\newblock {Constraining values of bag constant for strange star candidates}.
\newblock {\em Int. J. Mod. Phys. D} {\bf 2019}, {\em 28},~1941006.
\newblock {{https://doi.org/10.1142/S0218271819410062}}.

\bibitem[Benic \em{et~al.}(2015)Benic, Blaschke, Alvarez-Castillo, Fischer, and
Typel]{Benic:2014jia}
Benic, S.; Blaschke, D.; Alvarez-Castillo, D.E.; Fischer, T.; Typel, S.
\newblock {A new quark-hadron hybrid equation of state for astrophysics---I.
High-mass twin compact stars}.
\newblock {\em Astron. Astrophys.} {\bf 2015}, {\em 577},~A40.
\newblock {{https://doi.org/10.1051/0004-6361/201425318}}.

\bibitem[Yazdizadeh \em{et~al.}(2022)Yazdizadeh, Bordbar, and
Panah]{Yazdizadeh_2022}
Yazdizadeh, T.; Bordbar, G.; Panah, B.E.
\newblock The structure of hybrid neutron star in Einstein-$\lambda$ gravity
\newblock {\em Phys. Dark Universe} {\bf 2022}, {\em 35},~100982.
\newblock {{https://doi.org/10.1016/j.dark.2022.100982}}.

\bibitem[Eslam~Panah \em{et~al.}(2019)Eslam~Panah, Yazdizadeh, and
Bordbar]{EslamPanah:2018rfe}
Eslam~Panah, B.; Yazdizadeh, T.; Bordbar, G.H.
\newblock {Contraction of cold neutron star due to in the presence a quark
core}.
\newblock {\em Eur. Phys. J. C} {\bf 2019}, {\em 79},~815.
\newblock {{https://doi.org/10.1140/epjc/s10052-019-7331-1}}.

\bibitem[Jaikumar \em{et~al.}(2006)Jaikumar, Reddy, and
Steiner]{Jaikumar:2005ne}
Jaikumar, P.; Reddy, S.; Steiner, A.W.
\newblock {The Strange star surface: A Crust with nuggets}.
\newblock {\em Phys. Rev. Lett.} {\bf 2006}, {\em 96},~041101.
\newblock {{https://doi.org/10.1103/PhysRevLett.96.041101}}.

\bibitem[Ofek \em{et~al.}(2007)Ofek et~al.]{Ofek:2006vt}
Ofek, E.O.; Cameron, P.; Kasliwal, M.; Gal-Yam, A.; Rau, A.; Kulkarni, S.; Frail, D.; Chandra, P.; Cenko, S.; Soderberg, A.; et~al.
\newblock {SN 2006gy: An extremely luminous supernova in the early-type galaxy
NGC 1260}.
\newblock {\em Astrophys. J. Lett.} {\bf 2007}, {\em 659},~L13--L16.
\newblock {{https://doi.org/10.1086/516749}}.

\bibitem[{Ouyed} \em{et~al.}(2009){Ouyed}, {Leahy}, and
{Jaikumar}]{2009arXiv0911.5424O}
{Ouyed}, R.; {Leahy}, D.; {Jaikumar}, P.
\newblock {Predictions for signatures of the quark-nova in superluminous
supernovae}.
\newblock {\em arXiv} {\bf 2009}, arXiv:0911.5424.

\bibitem[Mukhopadhyay and Schaffner-Bielich(2016)]{Mukhopadhyay:2015xhs}
Mukhopadhyay, P.; Schaffner-Bielich, J.
\newblock {Quark stars admixed with dark matter}.
\newblock {\em Phys. Rev. D} {\bf 2016}, {\em 93},~083009.
\newblock {{https://doi.org/10.1103/PhysRevD.93.083009}}.

\bibitem[Panotopoulos and Lopes(2018)]{Panotopoulos:2018ipq}
Panotopoulos, G.; Lopes, I.
\newblock {Radial oscillations of strange quark stars admixed with fermionic
dark matter}.
\newblock {\em Phys. Rev. D} {\bf 2018}, {\em 98},~083001.
\newblock {{https://doi.org/10.1103/PhysRevD.98.083001}}.

\bibitem[Ruderman(1972)]{Ruderman:1972aj}
Ruderman, M.
\newblock {Pulsars: Structure and dynamics}.
\newblock {\em Ann. Rev. Astron. Astrophys.} {\bf 1972}, {\em 10},~427--476.
\newblock {{https://doi.org/10.1146/\linebreak annurev.aa.10.090172.002235}}.

\bibitem[Bowers and Liang(1974)]{Bowers:1974tgi}
Bowers, R.L.; Liang, E.P.T.
\newblock {Anisotropic Spheres in General Relativity}.
\newblock {\em Astrophys. J.} {\bf 1974}, {\em 188},~657--665.
\newblock {{https://doi.org/10.1086/\linebreak152760}}.

\bibitem[{Weber}(2017)]{2017paln.book.....W}
{Weber}, F.
\newblock {\em Pulsars as Astrophysical Laboratories for Nuclear and Particle
Physics}; Routledge: New York, NY, USA, 2017.

\bibitem[{Bordbar} and {Karami}(2022)]{2022EPJC...82...74B}
{Bordbar}, G.H.; {Karami}, M.
\newblock {Anisotropic magnetized neutron star}.
\newblock {\em Eur. Phys. J. C} {\bf 2022}, {\em 82},~74.
\newblock {{https://doi.org/10.1140/epjc/s10052-022-10038-0}}.

\bibitem[Sokolov(1998)]{Sokolov:1998svw}
Sokolov, A.I.
\newblock {Universal effective coupling constants for the generalized
Heisenberg model}.
\newblock {\em Fiz. Tverd. Tela} {\bf 1998}, {\em 40},~1284.

\bibitem[Sawyer(1972)]{Sawyer:1972cq}
Sawyer, R.F.
\newblock {Condensed pi- phase in neutron star matter}.
\newblock {\em Phys. Rev. Lett.} {\bf 1972}, {\em 29},~382--385.
\newblock {{https://doi.org/10.1103/\linebreak PhysRevLett.29.382}}.

\bibitem[{Herrera} and {Santos}(1997)]{1997PhR...286...53H}
{Herrera}, L.; {Santos}, N.O.
\newblock {Local anisotropy in self-gravitating systems}.
\newblock {\em Phys. Rep.} {\bf 1997}, {\em 286},~53--130.
\newblock {{https://doi.org/10.1016/S0370-1573(96)00042-7}}.

\bibitem[{Barreto} and {Rojas}(1992)]{1992Ap&SS.193..201B}
{Barreto}, W.; {Rojas}, S.
\newblock {An Equation of State for Radiating Dissipative Spheres in General
Relativity}.
\newblock {\em Astrophys. Space Sci.} {\bf 1992}, {\em 193},~201--215.
\newblock {{https://doi.org/10.1007/BF00643201}}.

\bibitem[{Letelier}(1980)]{1980PhRvD..22..807L}
{Letelier}, P.S.
\newblock {Anisotropic fluids with two-perfect-fluid components}.
\newblock {\em Phys. Rev. D} {\bf 1980}, {\em 22},~807--813.
\newblock {{https://doi.org/10.1103/\linebreak PhysRevD.22.807}}.

\bibitem[Kippenhahn \em{et~al.}(2012)Kippenhahn, Weigert, and
Weiss]{Kippenhahn:2012qhp}
Kippenhahn, R.; Weigert, A.; Weiss, A.
\newblock {\em {Stellar Structure and Evolution}}; 
Springer: Berlin/Heidelberg, Germany, 2012; ISBN~9783642303043.
\newblock {{https://doi.org/10.1007/978-3-642-30304-3}}.

\bibitem[Mak and T.(2002)]{MakANDHarko}
Mak, M.K.; Harko, T.
\newblock {An exact anisotropic quark star model}.
\newblock {\em Chin. J. Astron. Astrophys.} {\bf 2002}, {\em 2},~248--259.
\newblock {{https://doi.org/10.1088/\linebreak1009-9271/2/3/248}}.

\bibitem[Deb \em{et~al.}(2017)Deb, Chowdhury, Ray, Rahaman, and
Guha]{Deb:2016lvi}
Deb, D.; Chowdhury, S.R.; Ray, S.; Rahaman, F.; Guha, B.K.
\newblock {Relativistic model for anisotropic strange stars}.
\newblock {\em Ann. Phys.} {\bf 2017}, {\em 387},~239--252.
\newblock {{https://doi.org/10.1016/j.aop.2017.10.010}}.

\bibitem[Deb \em{et~al.}(2018)Deb, Roy~Chowdhury, Ray, and
Rahaman]{Deb:2015vda}
Deb, D.; Roy~Chowdhury, S.; Ray, S.; Rahaman, F.
\newblock {A New Model for Strange Stars}.
\newblock {\em Gen. Rel. Grav.} {\bf 2018}, {\em 50},~112.
\newblock {{https://doi.org/10.1007/s10714-018-2434-9}}.

\bibitem[Gabbanelli \em{et~al.}(2018)Gabbanelli, Rinc\'on, and
Rubio]{Gabbanelli:2018bhs}
Gabbanelli, L.; Rinc\'on, A.; Rubio, C.
\newblock {Gravitational decoupled anisotropies in compact stars}.
\newblock {\em Eur. Phys. J. C} {\bf 2018}, {\em 78},~370.
\newblock {{https://doi.org/10.1140/epjc/s10052-018-5865-2}}.

\bibitem[Ovalle(2017)]{Ovalle:2017fgl}
Ovalle, J.
\newblock {Decoupling gravitational sources in general relativity: From perfect
to anisotropic fluids}.
\newblock {\em Phys. Rev. D} {\bf 2017}, {\em 95},~104019.
\newblock {{https://doi.org/10.1103/PhysRevD.95.104019}}.

\bibitem[Ovalle \em{et~al.}(2018)Ovalle, Casadio, da~Rocha, and
Sotomayor]{Ovalle:2017wqi}
Ovalle, J.; Casadio, R.; da~Rocha, R.; Sotomayor, A.
\newblock {Anisotropic solutions by gravitational decoupling}.
\newblock {\em Eur. Phys. J. C} {\bf 2018}, {\em 78},~122.
\newblock {{https://doi.org/10.1140/epjc/s10052-018-5606-6}}.

\bibitem[Sharif and Butt(2018{\natexlab{a}})]{Sharif:2018pgq}
Sharif, M.; Butt, I.I.
\newblock {Complexity Factor for Charged Spherical System}.
\newblock {\em Eur. Phys. J. C} {\bf 2018}, {\em 78},~688.
\newblock {{https://doi.org/10.1140/epjc/\linebreak s10052-018-6121-5}}.

\bibitem[Sharif and Butt(2018{\natexlab{b}})]{Sharif:2018efi}
Sharif, M.; Butt, I.I.
\newblock {Complexity factor for static cylindrical system}.
\newblock {\em Eur. Phys. J. C} {\bf 2018}, {\em 78},~850.
\newblock {{https://doi.org/10.1140/epjc/\linebreak s10052-018-6330-y}}.

\bibitem[Abbas and Nazar(2018)]{Abbas:2018cha}
Abbas, G.; Nazar, H.
\newblock {Complexity Factor For Anisotropic Source in Non-minimal Coupling
Metric $f(R)$ Gravity}.
\newblock {\em Eur. Phys. J. C} {\bf 2018}, {\em 78},~957.
\newblock {{https://doi.org/10.1140/epjc/s10052-018-6430-8}}.

\bibitem[Herrera \em{et~al.}(2019)Herrera, Di~Prisco, and
Ospino]{Herrera:2019cbx}
Herrera, L.; Di~Prisco, A.; Ospino, J.
\newblock {Complexity factors for axially symmetric static sources}.
\newblock {\em Phys. Rev. D} {\bf 2019}, {\em 99},~044049.
\newblock {{https://doi.org/10.1103/PhysRevD.99.044049}}.

\bibitem[Prasad and Kumar(2022)]{Prasad:2021eju}
Prasad, A.K.; Kumar, J.
\newblock {Anisotropic relativistic fluid spheres with a linear equation of
state}.
\newblock {\em New Astron.} {\bf 2022}, {\em 95},~101815.
\newblock {{https://doi.org/10.1016/j.newast.2022.101815}}.

\bibitem[Herrera(2018)]{Herrera:2018bww}
Herrera, L.
\newblock {New definition of complexity for self-gravitating fluid
distributions: The spherically symmetric, static case}.
\newblock {\em Phys. Rev. D} {\bf 2018}, {\em 97},~044010.
\newblock {{https://doi.org/10.1103/PhysRevD.97.044010}}.

\bibitem[Sanudo and Pacheco(2009)]{Sanudo:2008bu}
Sanudo, J.; Pacheco, A.F.
\newblock {Complexity and white-dwarf structure}.
\newblock {\em Phys. Lett. A} {\bf 2009}, {\em 373},~807--810.
\newblock {{https://doi.org/10.1016/\linebreak j.physleta.2009.01.008}}.

\bibitem[Arias \em{et~al.}(2022)Arias, Contreras, Fuenmayor, and
Ramos]{Arias:2022qrm}
Arias, C.; Contreras, E.; Fuenmayor, E.; Ramos, A.
\newblock {Anisotropic star models in the context of vanishing complexity}.
\newblock {\em Ann. Phys.} {\bf 2022}, {\em 436},~168671.
\newblock {{https://doi.org/10.1016/j.aop.2021.168671}}.

\bibitem[Andrade and Contreras(2021)]{Andrade:2021flq}
Andrade, J.; Contreras, E.
\newblock {Stellar models with like-Tolman IV complexity factor}.
\newblock {\em Eur. Phys. J. C} {\bf 2021}, {\em 81},~889.
\newblock {{https://doi.org/10.1140/epjc/s10052-021-09695-4}}.

\bibitem{Contreras:2022vec}
E.~Contreras and Z.~Stuchlik,
Eur. Phys. J. C \textbf{82} (2022) no.8, 706
doi:10.1140/epjc/s10052-022-10684-4
[arXiv:2208.09028 [gr-qc]].

\bibitem{Bargueno:2022yob}
P.~Bargue\~no, E.~Fuenmayor and E.~Contreras,
Annals Phys. \textbf{443} (2022), 169012
doi:10.1016/j.aop.2022.169012
[arXiv:2208.09044 [gr-qc]].

\bibitem{Contreras:2021xkf}
E.~Contreras and E.~Fuenmayor,
Phys. Rev. D \textbf{103} (2021) no.12, 124065
doi:10.1103/PhysRevD.103.124065
[arXiv:2107.01140 [gr-qc]].

\bibitem[Gomez-Lobo(2008)]{Gomez-Lobo:2007mbg}
Gomez-Lobo, A.G.P.
\newblock {Dynamical laws of superenergy in General Relativity}.
\newblock {\em Class. Quant. Grav.} {\bf 2008}, {\em 25},~015006.
\newblock {{https://doi.org/10.1088/0264-9381/25/1/015006}}.

\bibitem[Herrera \em{et~al.}(2009)Herrera, Ospino, Di~Prisco, Fuenmayor, and
Troconis]{Herrera:2009zp}
Herrera, L.; Ospino, J.; Di~Prisco, A.; Fuenmayor, E.; Troconis, O.
\newblock {Structure and evolution of self-gravitating objects and the
orthogonal splitting of the Riemann tensor}.
\newblock {\em Phys. Rev. D} {\bf 2009}, {\em 79},~064025.
\newblock {{https://doi.org/10.1103/PhysRevD.79.064025}}.

\bibitem[Panotopoulos and Lopes(2018)]{Panotopoulos:2018joc}
Panotopoulos, G.; Lopes, I.
\newblock {Millisecond pulsars modeled as strange quark stars admixed with
condensed dark matter}.
\newblock {\em Int. J. Mod. Phys. D} {\bf 2018}, {\em 27},~1850093.
\newblock {{https://doi.org/10.1142/S0218271818500931}}.

\bibitem[Moraes \em{et~al.}(2021)Moraes, Panotopoulos, and
Lopes]{Moraes:2021lhh}
Moraes, P.H.R.S.; Panotopoulos, G.; Lopes, I.
\newblock {Anisotropic Dark Matter Stars}.
\newblock {\em Phys. Rev. D} {\bf 2021}, {\em 103},~084023.
\newblock {{https://doi.org/10.1103/PhysRevD.103.084023}}.

\bibitem[Panotopoulos and Rinc\'on(2019)]{Panotopoulos:2019wsy}
Panotopoulos, G.; Rinc\'on, A.
\newblock {Electrically charged strange quark stars with a non-linear
equation-of-state}.
\newblock {\em Eur. Phys. J. C} {\bf 2019}, {\em 79},~524.
\newblock {{https://doi.org/10.1140/epjc/s10052-019-7042-7}}.

\bibitem[Lopes \em{et~al.}(2019)Lopes, Panotopoulos, and
Rinc\'on]{Lopes:2019psm}
Lopes, I.; Panotopoulos, G.; Rinc\'on, A.
\newblock {Anisotropic strange quark stars with a non-linear
equation-of-state}.
\newblock {\em Eur. Phys. J. Plus} {\bf 2019}, {\em 134},~454.
\newblock {{https://doi.org/10.1140/epjp/i2019-12842-4}}.

\bibitem[Panotopoulos and Rinc\'on(2019)]{Panotopoulos:2019zxv}
Panotopoulos, G.; Rinc\'on, A.
\newblock {Relativistic strange quark stars in Lovelock gravity}.
\newblock {\em Eur. Phys. J. Plus} {\bf 2019}, {\em 134},~472.
\newblock {{https://doi.org/10.1140/epjp/i2019-12853-1}}.

\bibitem[Abell\'an \em{et~al.}(2020)Abell\'an, Rincon, Fuenmayor, and
Contreras]{Abellan:2020jjl}
Abell\'an, G.; Rincon, A.; Fuenmayor, E.; Contreras, E.
\newblock {Beyond classical anisotropy and a new look to relativistic stars: A
gravitational decoupling approach}
\emph{arXiv} {\bf 2020}, arXiv:2001.07961.
\newblock 

\bibitem[Panotopoulos \em{et~al.}(2020)Panotopoulos, Rinc\'on, and
Lopes]{Panotopoulos:2020zqa}
Panotopoulos, G.; Rinc\'on, A.; Lopes, I.
\newblock {Interior solutions of relativistic stars in the scale-dependent
scenario}.
\newblock {\em Eur. Phys. J. C} {\bf 2020}, {\em 80},~318.
\newblock {{https://doi.org/10.1140/epjc/s10052-020-7900-3}}.

\bibitem[Bhar \em{et~al.}(2020)Bhar, Tello-Ortiz, Rinc\'on, and
Gomez-Leyton]{Bhar:2020ukr}
Bhar, P.; Tello-Ortiz, F.; Rinc\'on, A.; Gomez-Leyton, Y.
\newblock {Study on anisotropic stars in the framework of Rastall gravity}.
\newblock {\em Astrophys. Space Sci.} {\bf 2020}, {\em 365},~145.
\newblock {{https://doi.org/10.1007/s10509-020-03859-6}}.

\bibitem[Panotopoulos \em{et~al.}(2020)Panotopoulos, Rinc\'on, and
Lopes]{Panotopoulos:2020kgl}
Panotopoulos, G.; Rinc\'on, A.; Lopes, I.
\newblock {Radial oscillations and tidal Love numbers of dark energy stars}.
\newblock {\em Eur. Phys. J. Plus} {\bf 2020}, {\em 135},~856.
\newblock {{https://doi.org/10.1140/epjp/s13360-020-00867-x}}.

\bibitem[Panotopoulos \em{et~al.}(2021{\natexlab{a}})Panotopoulos, Rinc\'on.
and Lopes]{Panotopoulos:2021obe}
Panotopoulos, G.; Rinc\'on, A.; Lopes, I.
\newblock {Interior solutions of relativistic stars with anisotropic matter in
scale-dependent gravity}.
\newblock {\em Eur. Phys. J. C} {\bf 2021}, {\em 81},~63.
\newblock {{https://doi.org/10.1140/epjc/s10052-021-08881-8}}.

\bibitem[Panotopoulos \em{et~al.}(2021{\natexlab{b}})Panotopoulos, Rinc\'on.
and Lopes]{Panotopoulos:2021dtu}
Panotopoulos, G.; Rinc\'on, A.; Lopes, I.
\newblock {Slowly rotating dark energy stars}.
\newblock {\em Phys. Dark Univ.} {\bf 2021}, {\em 34},~100885.
\newblock {{https://doi.org/10.1016/j.dark.2021.100885}}.

\bibitem[Becerra-Vergara \em{et~al.}(2019)Becerra-Vergara, Mojica.
Lora-Clavijo, and Cruz-Osorio]{Becerra-Vergara:2019uzm}
Becerra-Vergara, E.A.; Mojica, S.; Lora-Clavijo, F.D.; Cruz-Osorio, A.
\newblock {Anisotropic Quark Stars with an Interacting Quark Equation of
State}.
\newblock {\em Phys. Rev. D} {\bf 2019}, {\em 100},~103006.
\newblock {{https://doi.org/10.1103/PhysRevD.100.103006}}.

\bibitem[Panotopoulos \em{et~al.}(2022)Panotopoulos, Tangphati, and
Banerjee]{Panotopoulos:2021cxu}
Panotopoulos, G.; Tangphati, T.; Banerjee, A.
\newblock {Electrically charged compact stars with an interacting quark
equation of state}.
\newblock {\em Chin. J. Phys.} {\bf 2022}, {\em 77},~1682--1690.
\newblock {{https://doi.org/10.1016/j.cjph.2021.10.027}}.

\bibitem[Beringer \em{et~al.}(2012)Beringer et~al.]{ParticleDataGroup:2012pjm}
Beringer, J.; Arguin, J.F.; Barnett, R.M.; Copic, K.; Dahl, O.; Groom, D.E.; Lin, C.J.; Lys, J.; Murayama, H.; Wohl, C.G.; et~al.
\newblock {Review of Particle Physics (RPP)}.
\newblock {\em Phys. Rev. D} {\bf 2012}, {\em 86},~010001.
\newblock {{https://doi.org/10.1103/PhysRevD.86.010001}}.

\bibitem[Fiorella~Burgio and Fantina(2018)]{FiorellaBurgio:2018dga}
Fiorella~Burgio, G.; Fantina, A.F.
\newblock {Nuclear Equation of state for Compact Stars and Supernovae}.
\newblock {\em Astrophys. Space Sci. Libr.} {\bf 2018}, {\em 457},~255--335.
\newblock {{https://doi.org/10.1007/978-3-319-97616-7\_6}}.

\bibitem[Blaschke and Chamel(2018)]{Blaschke:2018mqw}
Blaschke, D.; Chamel, N.
\newblock {Phases of dense matter in compact stars}.
\newblock {\em Astrophys. Space Sci. Libr.} {\bf 2018}, {\em 457},~337--400.
\newblock {{https://doi.org/10.1007/978-3-319-97616-7\_7}}.

\bibitem[Silva \em{et~al.}(2015)Silva, Macedo, Berti, and
Crispino]{Silva:2014fca}
Silva, H.O.; Macedo, C.F.B.; Berti, E.; Crispino, L.C.B.
\newblock {Slowly rotating anisotropic neutron stars in general relativity and
scalar\textendash{}tensor theory}.
\newblock {\em Class. Quant. Grav.} {\bf 2015}, {\em 32},~145008.
\newblock {{https://doi.org/10.1088/0264-9381/32/14/145008}}.

\bibitem[Folomeev and Dzhunushaliev(2015)]{Folomeev:2015aua}
Folomeev, V.; Dzhunushaliev, V.
\newblock {Magnetic fields in anisotropic relativistic stars}.
\newblock {\em Phys. Rev. D} {\bf 2015}, {\em 91},~044040.
\newblock {{https://doi.org/10.1103/PhysRevD.91.044040}}.

\bibitem[Cattoen \em{et~al.}(2005)Cattoen, Faber, and Visser]{Cattoen:2005he}
Cattoen, C.; Faber, T.; Visser, M.
\newblock {Gravastars must have anisotropic pressures}.
\newblock {\em Class. Quant. Grav.} {\bf 2005}, {\em 22},~4189--4202.
\newblock {{https://doi.org/10.1088/0264-9381/22/20/002}}.

\bibitem[Horvat \em{et~al.}(2011)Horvat, Ilijic, and Marunovic]{Horvat:2010xf}
Horvat, D.; Ilijic, S.; Marunovic, A.
\newblock {Radial pulsations and stability of anisotropic stars with
quasi-local equation of state}.
\newblock {\em Class. Quant. Grav.} {\bf 2011}, {\em 28},~025009.
\newblock {{https://doi.org/10.1088/0264-9381/28/2/025009}}.

\bibitem[Arba\~nil and Panotopoulos(2022)]{Arbanil:2021ahh}
Arba\~nil, J.D.V.; Panotopoulos, G.
\newblock {Tidal deformability and radial oscillations of anisotropic
polytropic spheres}.
\newblock {\em Phys. Rev. D} {\bf 2022}, {\em 105},~024008.
\newblock {{https://doi.org/10.1103/PhysRevD.105.024008}}.

\end{thebibliography}
\end{document}